\documentclass[referee]{raa}
\usepackage{graphicx,times}
\usepackage{lscape}
\usepackage{natbib}
\usepackage{amssymb,amsmath}

\bibpunct{(}{)}{;}{a}{}{,}

\usepackage[a4paper=true,dvipdfm=true,pagebackref=true]{hyperref}
\hypersetup{pdftitle = The title of my PDF, pdfauthor = My name, pdfsubject= The subject, pdfkeywords = keyword1 keyword2 keyword3}
\hypersetup{colorlinks = true, linkcolor = green, anchorcolor = red, citecolor = blue, filecolor = red, pagecolor = red, urlcolor = red}

\begin{document}

\title{Astrophysical constraints on the proton-to-electron mass ratio with FAST}

 \volnopage{ {\bf 2012} Vol.\ {\bf X} No. {\bf XX}, 000--000}
   \setcounter{page}{1}

   \author{Xi Chen\inst{1,3}, Simon P. Ellingsen\inst{2}, Ying Mei\inst{1,}$^{\textbf{*}}$
   }

   \institute{Center for Astrophysics, GuangZhou University,
Guangzhou 510006, China; meiying@astrolab.cn \\
 \and
 School of Physical Sciences, University of Tasmania, Private Bag 37, Hobart, Tasmania 7001, Australia\\
 \and
   Shanghai Astronomical Observatory, Chinese Academy of
Sciences, Shanghai 200030, China \\
\vs \no
   {\small Received 2018 February 6; accepted 2018 October 16}
}

\abstract{That the laws of physics are the same at all times and
places throughout the Universe is one of the basic assumptions of
physics.  Astronomical observations provide the only means to test
this basic assumption on cosmological time and distance scales.
The possibility of variations in the dimensionless physical constant
 $\mu$ - the proton-to-electron mass ratio, can be tested by comparing
astronomical measurements of the rest frequency of certain spectral
lines at radio wavelengths with laboratory determinations. Different
types of molecular transitions have different dependencies on $\mu$
and so observations of two or more spectral lines towards the same
astronomical source can be used to test whether there is any
evidence for either temporal or spatial changes in the physical
fundamental constants.  $\mu$ will change if the relative strength
the strong nuclear force compared to the electromagnetic force
varies. Theoretical studies have shown that the rotational
transitions of some molecules which have transitions in the
frequency range which will be covered by the FAST (e.g., CH$_{3}$OH,
OH and CH) are sensitive to changes in $\mu$. A number of studies
looking for possible variations in $\mu$ have been undertaken with
existing telescopes, however, the greater sensitivity of FAST means
it will open new opportunities to significantly improve upon
measurements made to date. In this paper, we discuss which molecular
transitions, and sources (both in the Galaxy and external galaxies)
are likely targets for providing improved constraints on $\mu$ with
FAST.
  }

   \authorrunning{X. Chen, S. P. Ellingsen \& Y. Mei}            %author_head in even pages
   \titlerunning{Constraints on $\mu$ with FAST}  % title_head in odd pages
   \maketitle

%________________________________________________ sections below
%
\section{Introduction}           %% first-level sections will be auto-capitalized
\label{sect:intro}

Theories beyond the standard model of physics have predicted the
possibility of space-time variation in the fundamental constants.
Over the last few decades a number of laboratory studies,
theoretical calculations and astronomical observations have been
conducted to search for the signatures of such variations (for a
recent review of the field see \cite{Uzan11}. Astrophysical
spectroscopic studies have mostly focussed on searching for
variations in the fine structure constant $\alpha=e/\hbar c$, and
the proton-to-electron mass ratio $\mu=$m$_{p}$/m$_{e}$.
Astrophysical spectroscopy can be used to search for changes in the
dimensionless constants $\alpha$ and/or $\mu$ by comparing the rest
frequency of different transitions in atoms and molecules as a
function of time and/or position. The energy levels of different
quantum states can be expressed in terms which include the
dimensionless constants $\alpha$ and $\mu$. Where a transition takes
place between energy levels with different dependencies on these
constants a variation in either will cause a change of the
transition frequencies compared to the laboratory value (e.g., \cite{Rein06,Webb99}).
The change in frequency caused
by a change in either of these dimensionless constants is
characterized by the sensitivity coefficient of transition
$K_\alpha$ or $K_\mu$ (which is defined as the proportionality
constant between the fractional frequency shift of the transition,
$\Delta \nu/\nu$ and the fractional shift in $\alpha$ or $\mu$) as
follows:

\begin{equation}
  \frac{\Delta\nu}{\nu}=K_{\alpha}\frac{\Delta_{\alpha}}{\alpha}+K_{\mu}\frac{\Delta_{\mu}}{\mu}.
\label{eq:LebsequeI}
\end{equation}

The rest frequency of electron transitions in atoms is generally
more sensitive to the fine structure constant $\alpha$, whereas
rotational transitions in molecules have a stronger dependence on
the proton-to-electron mass ratio $\mu$. The mass of the proton is
set by the strong nuclear force, while the mass of the electron is
set by weak-electromagnetic force, any change in the relative
strength of these two fundamental interactions will change $\mu$ and
hence the rest frequency of rotational transitions. One mechanism
which may cause variability of the two constants is through the
scalar fields which are potential candidates for producing dark
energy, which is responsible for the observed cosmic acceleration.
The chameleon mechanism proposes that the scalar fields are
ultra-light in the cosmic vacuum but effectively possess a large
mass locally when they are coupled to ordinary matter \citep{Khoury04}. 
Therefore the searches for changes in $\mu$ and
$\alpha$ are especially important for not only understanding the
nature of the fundamental laws of physics, but also providing direct
observational tests for some cosmological theories.

Furthermore models of dynamical scalar fields predict relationships
between variations in different dimensionless constants such as
$\mu$ and $\alpha$ with time, for example:
\begin{equation}
  \frac{\dot{\mu}}{\mu}=R\frac{\dot{\alpha}}{\alpha},
\label{eq:LebsequeI}
\end{equation}
where $R$ is a scalar argued to be of order -40 to -50 \citep{Ave06,Tho13a,Tho13b}. Values of $R$ of this order imply that variations in $\mu$ will be easier to detect
than those in $\alpha$.  This means that upper limits on variations
in $\mu$ at relatively low redshift can significantly constrain
variations in $\alpha$ at higher redshift for example if rolling
scalar fields are the mechanism through which they are produced.

The most sensitive searches for possible spatial or temporal
variations of $\mu$ require high signal to noise observations of the
molecular transitions that have a large sensitivity coefficient
$K_\mu$. Previous searches for variations in $\mu$ on cosmological
timescales have compared optical transitions of molecular H$_{2}$
(the most abundant astrophysical molecule), in high-redshifted
objects with accurate laboratory measurements \citep{Rein06}. 
These observations show that $\Delta \mu/\mu < 3 \times
10^{-5}$ over look-back times of $\sim$12 Gyr. However, the H$_{2}$
transitions measured in these observations have relatively poor
$K_{\mu} $sensitivity coefficients in the range
-0.05$<K_{\mu}<$+0.02. Rotational transitions of molecules are
generally much more sensitive to changes in $\mu$ ($K_{\mu} \sim 1$)
than rovibrational transitions of H$_{2}$, therefore more recent
observational studies have focused on searching for changes in $\mu$
using rotational transitions of molecules at radio wavelengths. Some
molecular transitions have even greater sensitivity to changes in
$\mu$ than the majority, for example, the inversion transitions of
ammonia (NH$_{3}$) have $K_\mu$=-4.46.  Hence the ammonia inversion
transitions are $\sim$100 times more sensitive to variations in
$\mu$ than the H$_{2}$ transitions in the optical wavelength range.
Astronomical observations of NH$_{3}$ transitions at radio
wavelengths have been used to constrain $\Delta \mu/\mu <
4.7\times10^{-7}$ (1$\sigma$) in the $z=0.89$ lensing galaxy in the
PKS1830-211 gravitational lens system \citep{Hen09} and
$\Delta \mu/\mu < 1.2\times10^{-7}$ in the $z=0.68$ absorbing galaxy
towards the radio source B0218+357 \citep{Kane11}. In the nearby
universe observations of NH$_{3}$ transitions towards molecular
clouds in the Milky Way have been used to constrain variations in
$\mu$ to be $\Delta \mu/\mu<2\times10^{-8}$ ($3\sigma$; \cite{Levs13}). 
However these limits are achieved by comparing the
spectral profile of NH$_3$ inversion transitions with less sensitive
($K_{\mu} \sim 1$)  transitions from different species, such as
cyanopolynes in particular HC$_{3}$N transitions. This approach
suffers from difficult to quantify systematic effects due to the
unknown degree of chemical segregation between the different
molecular species (i.e. due to different and inhomogeneous spatial
distributions of the different molecules along the line of sight).

The major limitation on observations of molecular transitions at
radio and millimetre wavelengths is that most common astrophysical
molecules have the same, or very similar dependency on $\mu$, for
all of their transitions. A recent breakthrough has been the
discovery that the hindered internal rotation which produces the
rich microwave spectra observed in some polyatomic molecules (e.g.
methanol CH$_{3}$OH, and methyl mercaptan CH$_{3}$SH) also causes a
significant enhancement in $K_{\mu}$ compared to those observed in
rotational transitions in any other molecule commonly found in
interstellar space, including NH$_{3}$ \citep{Jan11a,Jan13,Levs11}. 
Furthermore, the different transitions of these molecules have different K$_{\mu}$
coefficients, meaning that they offer the opportunity to tightly
constrain $\mu$ using observations of a single molecular species,
thus avoiding chemical segregation issues that arise when comparing
transitions associated with different molecules. The high $K_{\mu}$
transitions are generally those between near degenerate levels of
these molecules, hence they are typically at low radio frequencies.
The low radio frequency range will be well covered by the
Five-Hundred-Meter Aperture Radio Telescope (FAST). FAST will open
new possibilities for making sensitive observations of weak
molecular emission from high-$z$ astronomical objects. In this
paper, we discuss the molecular transitions and sources (both in the
Galaxy and external galaxies) which are likely to provide the best
opportunity to make sensitive searches for possible variations in
$\mu$ with FAST.

\section{Performance of FAST}

FAST is a current Chinese mega-science project to build the largest
single dish radio telescope in the world. The telescope consists of
a 500-meter aperture with an illuminated aperture of 300-meters. The
telescope is located in the Guizhou province, China, and the first
phase covers a continuous frequency range, 70 MHz -- 3 GHz using a
set of 9 receivers (see \cite{Nan11,Li18}). The L-band
19-beam receiver is the main instrument for surveys of H{\sc i} and
pulsars in the Galaxy and nearby galaxies. The design specifies the
system temperature and resolution at L band to be $\sim$ 25 K and
3$'$, respectively. The declination range of FAST is -15$^\circ$ --
65$^\circ$. The combination of large collecting area and advanced
receiver and backend systems means that FAST will be an important
instrument for advancing our understanding of cosmology, galaxy
evolution, the interstellar medium life cycle, star formation and
exoplanets. A spectroscopic survey of Galactic and extragalactic
objects with continuous coverage between 70 MHz -- 3 GHz is one of
the main scientific programs which has been started with FAST \citep{Li13,Li18}. 
This frequency range includes a number of important
molecular transitions with different sensitivity to variations in
$\mu$ and these are discussed in Section 3.

\section{Molecular transition candidates for FAST}

Table 1 summarizes the molecular transitions which are sensitive to
variations in the proton-to-electron ratio and have rest frequencies
in the 70 MHz -- 3 GHz range which will be covered by FAST.

\subsection{CH$_{3}$OH and its isotopes}

Methanol (CH$_{3}$OH) is one of the simplest molecules that exhibits
hindered internal rotation and thus has been the subject of a number
of theoretical and observational studies relating to variations in
the proton to electron mass ratio (e.g.,\cite{Ell12,Jan11a,Jan11b,Levs11}). 
Methanol is a widespread
interstellar molecule observed in numerous regions in the Galaxy and
in some external galaxies (e.g., \cite{Her09, Mar06,Sjou10}). In the local
universe methanol emission is commonly observed in the vicinity of
high-mass star forming regions exhibiting both maser and thermal
emission from hot cores.  Absorption is also detected toward cold
clouds in the foreground of continuum sources. There are more than
30 methanol transitions known to exhibit maser emission with
wavelength in the range from centimeter to millimeter. These
transitions are empirically classified into two types which are
known as class I or class II transitions on the basis of the
locations where they are observed to arise in the star forming
region -- class I methanol masers usually arise from multiple
positions within a star forming region and are distribued on scales
of 0.1--1.0 parsec, whereas class II methanol masers are found
within $\sim1''$ of high-mass young stellar objects (e.g., \cite{Batrla88,Plam90}). 
Over one thousand methanol
maser sources have been detected in our Galaxy, including $\sim$ 900
class II (e.g. \cite{Green09}) and $\sim$ 400 class I methanol
maser sources (see the review of \citep{Chen14}). Observations of
both class I and class II methanol masers within the Milky Way have
recently been applied to constrain spatial variations in $\mu$ at
the level of $\Delta \mu/\mu < 3\times10^{-8}$ ($1\sigma$; \citep{Ell11} Levshakov
et al. 2011). However, at cosmologically
interesting distances there is only one detection of methanol and
that is in absorption towards PKS B1830-211. This system is a
gravitationally lensed quasar and the absorption occurs in the
lensing galaxy which is at a redshift of $z=0.89$.  Observations of
three different methanol transitions with rest frequencies of 12.2,
48.3 and 60.5 GHz have been used to constrain variations in $\mu$ on
temporal scales of around 7 Gyr (the look-back time to $z$ = 0.89)
\citep{Bagdo13a,Bagdo13b,Muller11} Ellingsen et al. 2012; 
The most sensitive of these observations by Bagdonaite et
al. constrain $\Delta \mu/\mu$ to be less than $1\times10^{-7}$
(2$\sigma$). An important point to note for FAST searches is that
the rest frequencies of the methanol transitions detected in the PKS
B1830-211 system far exceed the upper frequency limit of the first
phase of FAST (3 GHz). The class II and class I methanol masers with
the lowest rest frequencies are the $5_{1}-6_{0}$ $A^{+}$ transition
at 6.7 GHz and the $9_{-1}-8_{-2}$ $E$ transition at 9.9 GHz,
respectively.  These two transitions are also very sensitive to
variations in $\mu$ with K$_{\mu}$=-42 and 12, for the 6.7 and
9.9~GHz transitions, respectively. These transitions cannot be used
to look for variations in $\mu$ in the Milky Way or nearby galaxies
with FAST, however, where they may be present in galaxies at
redshifts of $>2$, they would be within the detectable frequency
range.

Theoretical calculations show that some lower-frequency methanol
transitions which have not been the target of previous searches for
variations in $\mu$ possess larger sensitivity coefficients than the
most commonly observed methanol maser transitions \citep{Jan11a,Jan11b}. We have collated a list of those transitions of methanol
and its isotopologues which lie within the FAST frequency range and
list them in Table 1. We also list the information for the lowest
rest frequency transitions for both class I and class II methanol
masers in this table, since they are potential candidates for
measuring temporal variations in $\mu$ through observations of
high-$z$ objects with FAST, we discuss this further in Section 4.
The equivalent transitions of methanol isotopologues are generally
more sensitive to changes in $\mu$ than those of methanol itself.
The most sensitive of the methanol isotopologues with $K_\mu=$330 is
the $1_{1}-2_{2}$ $E$ transition of $^{12}$CD$_{3}^{16}$OH which has
a rest frequency of approximately 1.2 GHz.  The sensitivity of this
transition to variations in $\mu$ is approximately one order of
magnitude larger than that of the highest $K_\mu$ methanol
transitions used in previous studies. This demonstrates that the
detection of methanol isotopologues would significantly help to make
more sensitive investigations for variations in $\mu$. Emission from
methanol isotopologues has been detected in both high-mass and
low-mass star forming regions in the Milky Way, for example
$^{13}$CH$_{3}$OH and CD$_{3}$OH have been detected by \cite{Parise02,Parise04} 
and \cite{Rat11}, showing that these
isotopologues are present at detectable abundances in the local
universe.  Although it should be noted that the detections of the
methanol isotopologues were from transitions at millimeter
wavelengths rather than the lower frequency transitions suitable for
observations with FAST.

It is worth noting that the sensitivity coefficients of the
transitions of methanol and its isotopologues listed in Table 1 have
both large positive and large negative values.  This means that a
variation in $\mu$ will shift the frequency of some transitions to
higher frequencies while others shift to lower frequencies. So the
most sensitive method for accurately probing for variations in $\mu$
is through simultaneous observations of different transitions with
large positive and negative values of $K_\mu$. Observations of
different isotopologues also have the advantage that they avoid many
systematic effects that can affect comparisons based on transitions
of different molecules (such as chemical segregation).

\subsection{Other molecules}

In addition to methanol, the frequency range of FAST will also cover
transitions of other important interstellar molecules which have good
sensitivity to variations in $\mu$.  These other molecules and the relevant
transitions are also listed in Table 1.

\textbf{CH} is abundant in the Universe and the two ground-state
$\Lambda$-doublet transitions for $^2\Pi_{3/2}$ $J=3/2$ and $J=1/2$
which have rest frequencies of $\sim0.7$ and $\sim3.3$ GHz,
respectively, have been observed towards numerous clouds in the
Milky Way (e.g., \citep{Genz79,Whit78,Ziu85}.  
The 3.3 GHz transition has also been detected in other
galaxies (e.g., \cite{Whit80}). Theoretical calculations show
that the two $\Lambda$-doublet transitions of CH are very sensitive
to changes in $\mu$ with $K_\mu$ ranging from 1.7 -- 6.3 \citep{Kozlov09}. 
Simultaneous astronomical observations of the two
$\Lambda$-doublet transitions of CH have been undertaken to
constrain $\mu$-variations at $1\sigma$ upper bounds of $\Delta
\mu/\mu<3\times10^{-7}$ in our Galaxy \citep{Truppe13}. It
should be noted that the frequency of $J=1/2$ transitions are
slightly above the 3 GHz upper limit of the first stage receivers
being developed for FAST. This means it will not be possible to
constrain variations in $\mu$ for Galactic objects with the two
$\Lambda$-doublet transitions of CH simultaneously with FAST.
However FAST will be suitable for simultaneously measuring both
transitions for moderate redshift objects (e.g. $z>0.1$). The higher
sensitivity of FAST will open the opportunity to detect relatively
weak emission from this molecule in high$-z$ objects.

\textbf{OH} is a very common interstellar molecule and has been
widely observed in our Galaxy and external galaxies. Maser emission
from the OH molecule has been observed from a number of transitions,
for example the ground-state transitions ($^{2}\Pi_{3/2}$, $J=3/2$
state), and many excited state-state transitions (including
$^{2}\Pi_{1/2}$, $J=1/2$ at 4765 MHz, and $^{2}\Pi_{3/2}$, $J=5/2$
at 6035 MHz). Of the various OH maser transitions the 1665/1667 MHz
ground-state transitions in star forming regions are usually the
strongest. At present about $\sim3000$ OH maser sources have been
detected in our Galaxy, most of them are stellar masers associated
with evolved stars (see \cite{mu10}).  At present only $\sim$ 400
OH masers have been detected in star forming regions (see \cite{Qiao14}), 
however, current sensitive OH maser surveys such as
SPLASH \citep{Daw14} and future surveys such as GASKAP \citep{Dickey13}
will significantly increase the number of OH
maser sources in the Galaxy (both evolved star and star forming
regions). In external galaxies, over one hundred galaxies with OH
megamaser activity have been found (e.g, \citep{Baan98,Darling02}). Similar to CH, the $\Lambda$-doublet
transitions of OH potentially also provide a very sensitive
indicator for searching for variations in $\mu$ \citep{Kozlov09}.
Observations of the ground-state 18cm OH lines in absorption at
$z=0.765$, have been used to constrain the variation in $\mu$ to be
$\Delta \mu/\mu<2.7\times10^{-6}$ for a look-back time of 6.7 Gyr
\citep{Kane12}. However, in that work the 21 cm hydrogen line
was adopted as a reference. Detection of more than one
$\Lambda$-doublet transition of OH offers the opportunity to further
improve the constraint. Within the frequency coverage of the first
stage of FAST, there are two OH $\Lambda$-doublet transitions at
$^2\Pi_{3/2}$ $J=3/2$ and $^2\Pi_{1/2}$ $J=9/2$ which may enable
such observations to be undertaken. The $^2\Pi_{1/2}$ $J=9/2$
transitions are very sensitive to changes in $\mu$ with the
sensitivity coefficient $K_\mu$ ranging from 210 -- 460, which is
more than two orders of magnitude greater than that of the 18 cm OH
lines used in previous studies. To date, the higher $J-$transitions
of OH in the 88 -- 192 MHz range have not been observed in
interstellar space. If these transitions can be detected in either
Galactic or extragalactic objects with FAST, it opens up the
possibility to make very sensitive studies for variations in $\mu$
using simultaneous observations of these OH transitions in
combination with ground-state OH. However, we note that these
higher$-J$ transitions are at significantly higher energies
\textbf{($E_{upper}=875$ K)} than the 18-cm lines, hence they may
have a different spatial distribution to the ground-state
transitions. This issue can be only clarified through detection and
observation of these transitions.

\textbf{CH$_{3}$NH$_{2}$} is a relatively small and stable molecule
which is abundant in the Milky Way (e.g., \cite{Lovas04}). It has
also been detected in a spiral galaxy at redshift $z=0.89$ (the
lensing galaxy in the 1830-211 system ; \cite{Muller11}).
CH$_{3}$NH$_{2}$ has hindered internal rotation of the CH$_{3}$
group with respect to the amino group (NH$_{2}$) which is similar to
what occurs in methanol.  In addition to this it also has tunneling
associated with wagging of the amino group. \cite{Ily12}
have used this molecule to search for temporal variations in $\mu$
through observations at millimeter wavelengths towards the $z=0.89$
intervening galaxy in the PKS 1830-211system. However, the
relatively low sensitivity $K_\mu$ for the transitions observed
means that it was not possible to place tight constraints on
variations in $\mu$ from those observations
($\Delta\mu/\mu<1\times10^{-5}$). Within the frequency coverage of
FAST, there are multiple CH$_{3}$NH$_{2}$ transitions, one of which
is very sensitive to variations in $\mu$ \citep{Ily12}. The
sensitivity of the various transitions have $K_\mu$ spanning the
range -1 -- -19 and observations in these transitions offer the
potential to make sensitive searches for variations in $\mu$ with
FAST.

\textbf{CH$_{3}$SH} is the sulphur analogue of methanol, therefore
similar to methanol it experiences hindered internal rotation which
results in larger sensitivity to variations in $\mu$. There are
multiple transitions of CH$_{3}$SH which lie within the frequency
coverage of FAST, and these have a large spread in their $K_\mu$
sensitivity coefficients which span an approximate range of -15 --12
\citep{Jan13}. It should be noted that the frequency of the
$4_{0}-3_{1}$ $A^+$ transition, which has the greatest sensitivity
to variations in $\mu$ ($K_\mu=-14.94)$ is above the 3 GHz upper
limit of the FAST frequency coverage, however, this transition will
be a candidate for observations in objects with moderate redshifts.
To date, CH$_{3}$SH has only been detected in the Milky Way (e.g.,\cite{Gibb00,Linke79}), however, the high sensitivity
of FAST will likely make it possible to detect this molecule in some
external galaxies.

\textbf{C$_{2}$H$_{6}$O$_{2}$} has recently been shown to have
low-frequency transitions (within the FAST frequency coverage) which
are sensitive to variations in $\mu$, with $K_\mu$ ranging from -17
(for the 882.2 MHz transition) to 18 (for the 978.3 MHz transition).
\cite{Viat14} have calculated the $K_\mu$ sensitivity
coefficient for approximately 10 transitions which lie within the
FAST frequency range (see Table 3 of \cite{Viat14}). Here we
list only those transitions with larger $K_\mu$ values in Table 1.
This molecule has been detected in interstellar space in the comet
C/1995 O1 (Hale-Bopp; \citep{crovi04}) and in molecular clouds
in the center of the Milky Way \citep{Hollis02}, although the
transitions observed to date are from higher frequency transitions
beyond the upper limit of the FAST frequency coverage.  It is very
likely that the high sensitivity of FAST will enable the detection
of the lower frequency transitions of interest here, both in the
Milky Way and perhaps also in some extragalactic sources.

\section{Astronomical targets for FAST}

In this section, we discuss which sources in our Galaxy and other
galaxies are prime targets for FAST observations to search for
possible spatial and temporal variations in $\mu$. We mainly focus
on targets for the methanol and OH transitions because they possess
the largest sensitivity coefficients to variations in $\mu$ (one to
two orders of magnitude more sensitive than most molecular
transitions) and are widespread throughout the Galaxy and external
galaxies, compared to some of the other less abundant molecules
discussed in Section 3.

\subsection {Targets in our Galaxy}

Sources which exhibit maser emission (including OH, CH$_{3}$OH and
H$_{2}$O) may provide the best target samples for searches for
possible variations in $\mu$ in the Milky Way. The strongest maser
emission  is usually observed from the molecular gas associated with
massive star forming regions. These clouds consist of gas which
contains relatively high abundances of both simple and complex
molecular species, including the molecules which have the greatest
sensitivity to variations in $\mu$ such as OH and methanol.

Observations focusing on transitions of the methanol isotopologues
(such as CH$_{3}$OD, as discussed in Section 3.1) can potentially
provide the most sensitive tests for variations in $\mu$. For a
source to exhibit methanol maser emission it must have a relatively
high abundance of the methanol molecule, therefore such regions are
likely to provide the best targets for detections of the methanol
isotopologues. There are over one thousand methanol maser sources
(including class I and class II transitions) which have been
detected in our Galaxy. In addition to these known methanol maser
detections from the past surveys, e.g. Parkes methanol multi-beam
survey at 6.7 GHz class II transition \citep{Green09}, a number
of new methanol maser surveys in our Galaxy are underway or
proposed. In particular, a series of targeted surveys for class I
methanol masers at 95 GHz transition have detected about 200 new
class I methanol maser sources, and combined with previous
observations they have increased the number of known class I
methanol masers in our Galaxy to $\sim$400 \citep{chen11,chen12,chen13,Gan13}. 
Statistical analysis of these surveys has
been used to predict that our Galaxy may contain at least $\sim$
2000 class I methanol maser sources, and suggested that the Bolocam
Galactic Plane Survey (BGPS) 1.1 mm dust continuum sources provide
more reliable samples for targeting further class I methanol maser
searches - about 1000 class I methanol masers are expected to be
detected from BGPS catalog targets. A new large survey for class I
methanol masers towards BGPS-selected sources is currently underway
with the Purple Mountain Observatory (PMO) 13.7-m radio telescope to
significantly increase the sample of class I methanol masers known
in the Galaxy. The majority of previous methanol maser surveys in
both the class I and class II transitions have been undertaken in
the southern hemisphere. Therefore, increasing the known number of
methanol maser sources in the northern sky is especially important
as this is the region accessible for FAST observations.  The MALT-45
survey being undertaken with using the Australia Telescope Compact
Array as six single-dishes \citep{Jor13} has made a complete
survey of 10 square degrees of the southern Galactic plane at 44
GHz.  This survey is the first sensitive, complete survey for any
class I methanol maser transition and will allow  for a much more
accurate estimate of the total number of class I methanol maser
sources throughout the Galaxy.

Furthermore, the newly-built Shanghai 65 m telescope will be used in
new searches for 6.7 GHz class II methanol and ground-state OH
masers, towards newly-identified samples of young stellar objects
from a number surveys of the Galactic Plane at mid-infrared
wavelengths, including the Spitzer Galactic Legacy Infrared
Mid-Plane Survey Extraordinaire (GLIMPSE). These searches will
increase the number of target sources for FAST $\mu$ variation
observations in both OH and methanol transitions in the northern
hemisphere.

\subsection {Targets in other galaxies}

The prime extragalactic targets to search for variations in $\mu$
are those which host OH megamaser emission, as observations of
multiple OH transitions can provide strong constraints. OH megamaser
emission has been detected in over 100 galaxies in surveys
undertaken to date (e.g, \cite{Baan98,Darling02}), 
however, most of the detected OH megamaser galaxies are at
relatively low redshifts with $z<0.2$. A survey for OH megamasers in
high-$z$ galaxies is required to place the most stringent
constraints on variations in $\mu$ over the history of the Universe.
OH megamaser searches in high-$z$ galaxies are also one of the main
early science projects for FAST (see \cite{Zhang12} and \cite{Li13}). 
Moreover, simultaneous H{\sc i} observations towards OH
megamaser galaxies with FAST also provide opportunities to constrain
variations in $\mu$, although the systemic effects due to comparing
observations from different molecules are difficult to quantify and
remove, as it is never really possible to determine the extent to
which they arise from different locations and to which the observed
spectral differences are due to different line of sight motions from
the two transitions.

There are approximately 10 extragalactic detections of CH emission
\citep{Bottine91,Whit80} and these represent
potential targets for searching for variations in $\mu$.  They are
however, all relatively nearby galaxies and therefore it will not be
possible to simultaneously observe the two CH transitions listed in
Table 1 with FAST. The galaxies towards which CH emission has been
detected often also exhibit strong OH absorption and some sometimes
H$_{2}$O emission. Therefore moderate redshift galaxies ($z>$0.1)
with OH absorption/emission or H$_{2}$O emission are potential
targets for future CH observations with FAST. In particular,
H$_{2}$O megamaser galaxies may be good targets because they can be
detected at relatively high redshifts, with the most distant source
being at $z=2.64$ (MG J0414+0534; \cite{Imp08b}).

Targets suitable for using methanol transitions to investigate
possible changes in $\mu$ are at present limited to one
extragalactic source outside the nearby galaxies. The lensing galaxy
in the PKS B1830-211 gravitational lense system is at a modest
redshift of $z=0.89$ and absorption from three different methanol
transitions has been detected towards it \citep{Bagdo13a,Bagdo13b,Ell12,Muller11}.
Emission from a
number of thermal methanol transitions at millimeter wavelengths
(mainly the $2_{k}-1_{k} E$ series at 96.7 GHz) has been detected
towards a handful of nearby galaxies (e.g.,\citep{Hen87,Hutt97}). 
However the detected emission from
these thermal lines is weak and broad, and hence is unlikely to be
able to provide useful constraints or tests for variations in $\mu$
(see \citep{Ell11}). While the absorption of methanol may
provide a more efficient approach for such constraints since its
spectrum is usually narrow, moreover besides the PKS B1830-211
discussed above, methanol absorption is also detected in the nearby
galaxy NGC3079 \citep{Imp08a}, suggesting that methanol
absorption may be common in external galaxies. More sensitive
searches for methanol emission (preferably maser emission) or
absorption in nearby extragalactic sources is required to better
understand how the properties of these sources are influenced by
their environment (factors such as metallicity and uv flux). A
number of searches have been made for class II methanol megamasers
at 6.7 and 12.2 GHz transitions towards samples selected from OH and
H$_{2}$O megamaser galaxies, and/or (Ultra-) Luminous Infrared
Galaxies ([U]LIRGs) \citep{Darling03,Ell94,Norr87,Phi98}. To date more than
one hundred sources have been searched, without any detections. It
may be, as suggested by \cite{Phi98}, that the mechanisms
which produce high methanol abundance in individual star formation
regions may not operate with sufficient efficiency on the larger
scales needed to produce class II megamasers. Alternatively, it may
be that the sensitivity of previous searches was not high enough to
detect maser emission in these sources or that appropriate targeting
criteria have not been identified to search for class II methanol
megamasers. A sensitive survey for class II methanol megamasers
towards a large sample of sources with FAST would clarify these
issues. Although for the first stage of FAST the targets for 6.7 GHz
class II methanol megamaser searches must be at redshifts of
$z>1.3$, in order for the emission to be detectable in the frequency
range of the telescope.

In contrast to class II methanol transitions, theoretical models
suggest that it is perhaps more likely that class I methanol
transitions can be excited on large scales in the central regions of
luminous galaxies \citep{Sobb93}. The recent detection of widespread
methanol maser emission in the 36 GHz class I transition toward the
center of the Milky Way \citep{Yuse13}, further supports
the theory. Based on these results, the first sensitive survey for
class I methanol megamasers in the 36 GHz transition was undertaken
towards a sample of OH megamaser galaxies with the Australia
Telescope Compact Array, and has produced the first extragalactic
detections of this transition towards NGC\,253 \citep{chen18,Ell14,Ell17}, Arp\,220 \citep{chen15},
NGC\,4945 \citep{McC17}, IC\,342 and NGC\,6946 \citep{Gorski18}. Further comparison with the infrared, radio and molecular
emission, and star formation rates of the host galaxies with/without
detections will enable us to refine targeting criteria for future
observations. Then by compiling a reliable target sample based on
these criteria, the 9.9 GHz class I methanol transition can be
searched towards galaxies meeting these criteria at redshifts
$z>2.3$ with FAST. The detection of two or more methanol megamaser
transitions in high-$z$ galaxies offers the best current prospects
for making sensitive observations for variations in $\mu$ at larger
look-back times.

\section{Summary}

Thanks to the large collecting area and advanced receiver and
backend systems, FAST should become one of the most important
instruments for searching for possible variations in the
proton-to-electron mass ratio $\mu$ on cosmological time and
distance scales. Within the frequency coverage of FAST, there are a
number of transitions of abundant interstellar molecules (e.g.,
CH$_{3}$OH, OH and CH) which are $1-2$ orders of magnitude more
sensitive to variations in $\mu$ than those typically used in
current studies. Existing and ongoing surveys for methanol and OH
masers in our Galaxy appear to provide the best samples for
observations to determine if $\mu$ varies spatially in the local
universe. However, further surveys for OH, CH and methanol
megamasers in high-$z$ galaxies are required to provide good quality
targets for investigations of variations in $\mu$ on cosmological
time scales. It is our hope that the potential importance of these
molecular megamasers for testing variations in $\mu$ will stimulate
broader surveys for these sources with FAST. FAST observations
utilising the most sensitive molecular transitions are likely to
improve the evidence either for or against the presence of
variations in the proton-to-electron mass ratio by more than $1-2$
orders of magnitude beyond the best current limits.

\normalem
\begin{acknowledgements}

This work was supported by the National Natural Science Foundation
of China (11590781), the Strategic Priority Research Program of the
Chinese Academy of Sciences (CAS; Grant No. XDA04060701), Key
Laboratory for Radio Astronomy, CAS.

\end{acknowledgements}

\clearpage

\begin{table}[]
\begin{center}
\begin{minipage}[]{200mm}
\caption[]{Selected molecular transitions and $K_\mu$ coefficients
for the FAST search\label{tab1}}\end{minipage}
\setlength{\tabcolsep}{0.1in} \small
 \begin{tabular}{crccr}
  \hline \hline \noalign{\smallskip}
Molecules & Transitions & Rest Frequency (MHz) & $K_\mu$ & Ref.\\
  \hline
$^{12}$CH$_{3}^{16}$OH & $1_{1}-1_{1}$ $A^\mp$ &  834.280          & -1.03    &  \cite{Jan11b}         \\
                       & $5_{1}-6_{0}$ $A^+$$^a$   & 6 668.567     & -42.0    &  \cite{Jan11a,Jan11b}      \\
                       & $9_{-1}-8_{-2}$ $E$$^a$   & 9 936.137     &  11.5    &   \cite{Jan11a,Jan11b}       \\
$^{12}$CD$_{3}^{16}$OH & $1_{1}-2_{2}$ $E$     & 1 202.296         & 330.0    &  \cite{Jan11b}     \\
                       & $2_{0}-1_{1}$ $E$     & 1 424.219         & -42.0    &  \cite{Jan11b}   \\
                       & $9_{2}-8_{3}$ $E$     & 2 827.262         &  43.0    &  \cite{Jan11b}   \\
                       & $5_{1}-6_{0}$ $A^+$   & 2 971.067         &  93.0    &  \cite{Jan11b}   \\
$^{12}$CH$_{3}^{18}$OH & $9_{2}-10_{1}$ $A^-$  & 2 604.912         &  93.0    &  \cite{Jan11b} \\
$^{12}$CD$_{3}^{16}$OD & $1_{0}-1_{-1}$ $E$    & 2 237.883         &  45.0    &  \cite{Jan11b} \\
                       & $7_{-4}-8_{-3}$ $E$   & 2 329.088         & -80.0    &  \cite{Jan11b}  \\
$^{13}$CH$_{3}^{16}$OH & $9_{-1}-8_{-2}$ $E$   & 1 989.502         & -63.0    &  \cite{Jan11b}  \\
\cline{1-5}
CH & $^2\Pi_{3/2}$ $J=3/2$ $F=2-2$             &   701.68          &  6.15    & \cite{Kozlov09}    \\
                       & $F=1-2$               &   703.97          &  6.32    &  \cite{Kozlov09}      \\
                       & $F=2-1$               &   722.30          &  6.17    &   \cite{Kozlov09}    \\
                       & $F=1-1$               &   724.79          &  5.97    &   \cite{Kozlov09}    \\
  & $^2\Pi_{1/2}$ $J=1/2$ $F=0-1$$^a$          & 3 263.795         &  1.71    &   \cite{Kozlov09}    \\
                       & $F=1-1$$^a$           & 3 335.481         &  1.70    &   \cite{Kozlov09}    \\
                       & $F=1-0$$^a$           & 3 349.194         &  1.69    &   \cite{Kozlov09}    \\
\cline{1-5}
OH & $^2\Pi_{3/2}$ $J=3/2$ $F=1-2$             & 1 612.231         &  2.61    &   \cite{Kozlov09}     \\
                       & $F=1-1$               & 1 665.402         &  2.55    &   \cite{Kozlov09}    \\
                       & $F=2-2$               & 1 667.360         &  2.55    &   \cite{Kozlov09}     \\
                       & $F=2-1$               & 1 720.530         &  2.49    &   \cite{Kozlov09}    \\
  & $^2\Pi_{1/2}$ $J=9/2$ $F=5-4$              &    88.950         &  459.9   &  \cite{Kozlov09}     \\
                       & $F=5-5$               &   117.150         &  349.59  &   \cite{Kozlov09}    \\
                       & $F=4-4$               &   164.796         &  248.77  &   \cite{Kozlov09}    \\
                       & $F=4-5$               &   192.996         &  212.68  &   \cite{Kozlov09}     \\
\cline{1-5}

CH$_{3}$NH$_{2}$       & $1(1) A_2-1(1) A_1$   &   879.859         &  -1.0    &   \cite{Ily12}  \\
                       & $1(1) B_2-1(1) B_1$   &   881.386         &  -1.0    &   \cite{Ily12} \\
                       & $1(1) A_2-2(0) A_1$   & 2 166.305         &  -19.1   &   \cite{Ily12} \\
                       & $2(1) A_1-2(1) A_2$   & 2 639.491         &  -1.0    &   \cite{Ily12} \\
                       & $2(1) B_1-2(1) B_2$   & 2 644.073         &  -1.0    &   \cite{Ily12} \\
\cline{1-5}

CH$_{3}$SH             & $1_{1}-1_{1}$ $A^\mp$ &   523.147         &  -1.0    &   \cite{Jan13} \\
                       & $2_{1}-2_{1}$ $A^\mp$ & 1 569.410         &  -1.0    &   \cite{Jan13}  \\
                      & $3_{1}-4_{0}$ $E$     & 1 874.635         &  11.77   &   \cite{Jan13}  \\
                       & $4_{0}-3_{1}$ $A^+$$^a$   & 3 038.566     & -14.94   &   \cite{Jan13}  \\
\cline{1-5}

C$_{2}$H$_{6}$O$_{2}$  & $1(1,1) v=0-1(0,1) v=1$&  2828.6           & -6.3    & \cite{Viat14} \\
                       & $2(1,2) v=0-2(0,2) v=1$&  1957.9           & -9.3    &   \cite{Viat14} \\
                       & $3(1,3) v=0-3(0,3) v=1$&   882.2           & -16.5   &   \cite{Viat14} \\
                       & $4(0,4) v=1-4(1,4) v=0$&   978.3           &  17.8   &   \cite{Viat14} \\
                       & $4(1,3) v=0-4(1,4) v=1$&   978.3           & -6.2    &  \cite{Viat14} \\

 \noalign{\smallskip}\hline
\end{tabular}
\end{center}
\tablecomments{0.86\textwidth}{The frequency of these transitions is
beyond the upper limit of 3 GHz of FAST frequency coverage. These
transitions can be only used to constraint on $\mu$-variation for
the targets in moderate redshifts.}
\end{table}

\end{document}